# Squash: A Scalable Quantum Mapper Considering Ancilla Sharing

Mohammad Javad Dousti, Alireza Shafaei, and Massoud Pedram
Department of Electrical Engineering, University of Southern California, Los Angeles, CA, U.S.A.
{dousti, shafaeib, and pedram}@usc.edu

## ABSTRACT

Quantum algorithms for solving problems of interesting size often result in circuits with a very large number of qubits and quantum gates. Fortunately, these algorithms also tend to contain a small number of repetitively-used quantum kernels. Identifying the quantum logic blocks that implement such quantum kernels is critical to the complexity management for realizing the corresponding quantum circuit. Moreover, quantum computation requires some type of quantum error correction coding to combat decoherence, which in turn results in a large number of ancilla qubits in the circuit. Sharing the ancilla qubits among quantum operations (even though this sharing can increase the overall circuit latency) is important in order to curb the resource demand of the quantum algorithm. This paper presents a multi-core *reconfigurable quantum processor* architecture, called *Requp,* which supports a layered approach to mapping a quantum algorithm and ancilla sharing. More precisely, a scalable quantum mapper, called *Squash,* is introduced, which divides a given quantum circuit into a number of quantum kernels—each kernel comprises *k* parts such that each part will run on exactly one of *k* available cores. Experimental results demonstrate that Squash can handle large-scale quantum algorithms while providing an effective mechanism for sharing ancilla qubits.

## 1. INTRODUCTION

Mapping quantum circuits directly to a quantum fabric is a challenging task due to the gigantic size of quantum circuits. These circuits comprise of two parts: a netlist of *quantum logical operations* followed by the *quantum error correction* (QEC) circuit. The QEC increases the circuit size by one or two orders of magnitude depending on the decoherence degree and the desired fidelity of results. To handle this growth in the size, circuits are mapped in two levels. The lower-level mapping, which is done by the *physical-level mapper*, maps a universal set of quantum operations in a fault-tolerant fashion followed by an appropriate QEC circuit to a given *physical machine description* (PMD). In the higher-level mapping, which is performed by the *logical-level mapper*, the *logical* circuit is mapped to an abstraction of the PMD assuming that the universal set of fault-tolerant quantum operations is provided by the lower level. This approach addresses the increase in size by the QEC in the first level very well, but it does not help for the second level. Real-size quantum circuits (even without QEC) are so large that traditional mappers introduced by previous researchers cannot efficiently handle them [1].

Reference [2] shows that Shor's factorization algorithm for a 1024-bit integer has $1.35\times10^{15}$ physical instructions. Assuming that the one-level ⟦7,1,3⟧ Steane code is used in this implementation, each logical operation results in about $10^5$ physical instructions. Hence, this algorithm has almost $1.35\times10^{10}$ logical operations. As can be seen, the physical-level mapper can handle the low-level QEC in a reasonable time as the number of physical instructions is not so high (~$10^5$ physical instructions). On the other hand, mapping $1.35\times10^{10}$ logical operations is very time consuming.

Fortunately, quantum circuits can be partitioned into multiple quantum computational stages. These stages tend to contain a small number of repetitively-used quantum kernels. This means that mapping one instance of these kernels is sufficient. For instance, Figure 1 shows the phase estimation algorithm which is the core of many well-known and useful quantum algorithms such as Shor's factorization algorithm [3] and quantum random walk [4]. As can be seen, in this circuit the controlled unitary is a kernel which is repeated $n$ times throughout the circuit with different exponents (throughout stages 2 to $n+1$). The exponent denotes the number of repetitions for the corresponding circuit. Clearly, identifying the quantum kernels and avoiding the remapping can exponentially improve the mapping speed for this circuit.

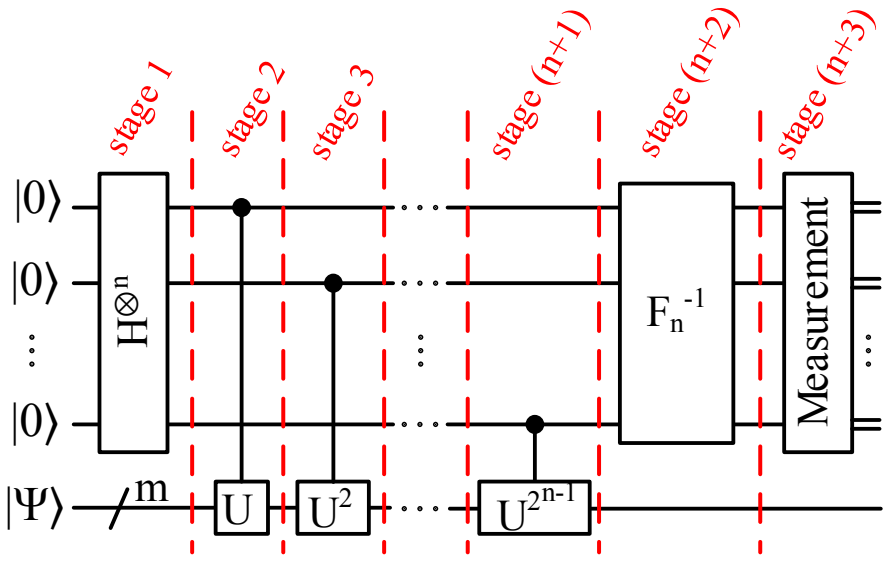

**Figure 1. Quantum circuit representation of the phase estimation algorithm [3]. The computational stages are identified in this circuit.**

Another major stumbling block for realizing a scalable quantum computer is the limited amount of physical qubits. Each logical operation is implemented in a fault-tolerant manner based on the adopted QEC code, and using a certain amount of *physical data qubits* and *physical ancilla qubits*. Physical data qubits are uniquely belong to their corresponding logical qubits, and hence cannot be shared. However, physical ancilla qubits, which are used to store intermediate information, may participate in the QEC circuit of various logical operations at different time instances. This reuse of ancilla qubits is referred to as *ancilla sharing*. Escalating the ancilla sharing increases the latency of the entire circuit while saving the precious quantum resources and vice versa. This trade-off is similar to the well-known area-delay trade-off in the VLSI circuits.

This paper introduces a novel quantum architecture, called *reconfigurable quantum processor architecture* (Requp), in order to address the problem of ancilla sharing. Requp has $k$ quantum cores each of which contains a *quantum reconfigurable compute region* (QRCR), a dedicated quantum cache, and a quantum memory. Quantum cores are arranged on a 2-D mesh topology. Each QRCR has a constrained amount of ancilla qubits while trying to share this limited resource among several quantum operations so as to minimize the latency. The major contribution of this architecture lies in its reconfigurability where it supports quantum operations with different number of ancilla qubits. This difference is quite substantial and neglecting it leads to over provisioning of quantum physical qubits.

Using the kernel extraction method and the proposed architecture (Requp) mentioned above, a scalable quantum mapper, called *scalable quantum mapper considering ancilla sharing* (*Squash*), is introduced. Squash initially divides the given circuit into a number of quantum kernels. For each kernel, it builds a *quantum operation dependency graph* (QODG) based on the data dependency among the operations. QODG is then partitioned into $k$ sub-graphs and bound to the quantum cores. These sub-graphs are subsequently scheduled and mapped to the Requp with $k$ quantum cores. Finally, results of mapping for each quantum kernel are combined in order to generate the entire mapping of the given circuit.

The rest of this paper is organized as follows: Section 2 explains the basics of quantum computing as well as the related work. Section 3 presents the new architecture (Requp), whereas Section 4 explains the proposed mapper (Squash). Experimental results are presented in Section 5, and finally Section 6 concludes the paper.

# 2. PRELIMINARIES AND PRIOR WORK

## 2.1 Quantum Computing Basics

A quantum bit, *qubit*, is a physical object (e.g., an ion or a photon) that carries data in quantum circuits. Qubits interact with each other through *quantum gates*. Depending on the underlying quantum computing technology, a universal set of quantum gates is available at the physical level. More precisely, each quantum fabric is natively capable of performing a universal set of one and two-qubit instructions (also called *physical instructions*). However, the importance of fault-tolerant quantum computation dictates the quantum circuits to be generated from *fault-tolerant* (FT) quantum operations. A universal (but redundant) set of FT operations includes H, S, T, $T^\dagger$, X, Y, Z, and CNOT operations [3], which may differ from physical instructions supported at the physical level. Fortunately, each FT quantum operation (or quantum operation for short) can be realized by using a composition of these physical instructions. Accordingly, a logical level circuit contains quantum operations where QEC is also applied.

A quantum circuit fabric is arranged as a 2-D array of identical *cells*. Each cell contains sites for creating qubits, reading them out, performing instructions on one or two physical qubits, and resources for routing qubits (or equivalently swapping their information to the neighboring qubit). In practice, however, an abstract *quantum architecture* (QA) is built which hides the physical information and the QEC details. Operation sites in this QA are capable of performing any quantum operation. The QA is also equipped with syndrome extraction circuitries following the quantum operation in order to prevent error propagation that may have been introduced by the quantum operation.

A *quantum compilation/synthesis* tool generates a reversible quantum circuit composed of quantum operations. Every qubit in the output circuit is called a *logical* qubit, which is subsequently encoded into several *physical* qubits in order to detect and correct potential errors on qubits. Physical qubits are comprised of two types: 1) *physical data qubits* and 2) *physical ancilla qubits*. Physical data qubits carry the encoded data of the logical qubits. Based on the type and the concatenation level of the QEC, a logical qubit is encoded to seven or more physical data qubits. On the other hand, physical ancilla qubits are used as scratchpads and can be *shared* among different logical qubits for the error correction procedure.

A *high-level mapping* tool schedules, places, and routes the logical circuit on the QA. To achieve this, the quantum algorithm is initially modeled as a *quantum operation dependency graph* (QODG), in which nodes represent quantum operations and edges capture data dependencies [1]. More precisely, operation $O_j$ depends on operation $O_i$ if $O_i$ and $O_j$ share at least one qubit and $O_j$ is the first operation after $O_i$ in the circuit that uses this (these) shared qubit(s). This dependency is shown as $O_i \rightarrow O_j$. For instance, Figure 2 depicts an FT implementation of a 3-input Toffoli operation [5] along with its QODG.

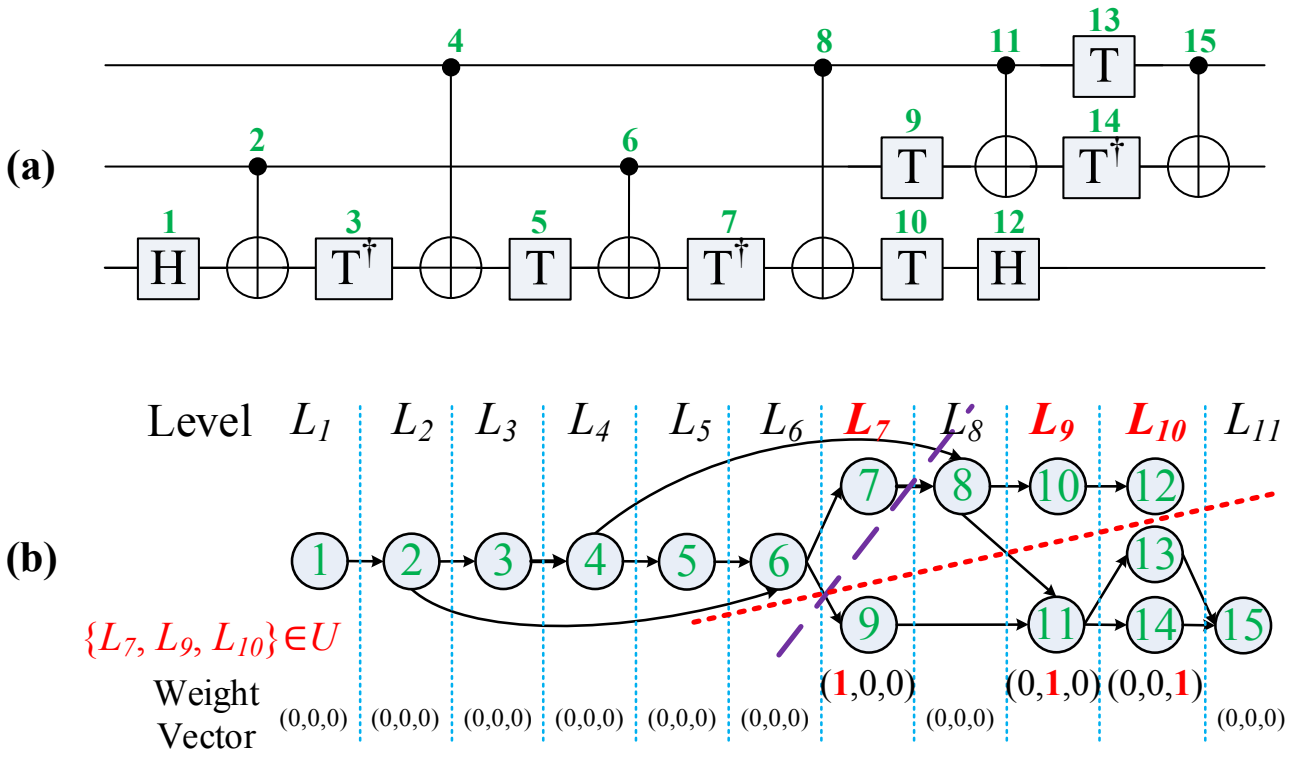

**Figure 2. (a) FT implementation of a three-input Toffoli circuit [5], (b) the corresponding QODG where each node represents a circuit operation. Detailed steps of the 2-way partitioning algorithm are also illustrated.**

Next, the QODG is mapped to the desired QA. The latency of the quantum algorithm mapped to the QA can be calculated as the length of the longest path (critical path) in the mapped QODG, where the length of a path in the QODG is in turn the summation of latencies of operations located on that path plus routing latencies of their qubit operands [1]. Critical path of the mapped QODG may not be the same as the original QODG, since the latter does not contain routing latencies. This can change the scheduling slacks, and hence may increase the critical path of the entire graph.

## 2.2 Prior Work

• **Quantum Architectures.** Metodi *et al.* propose the first QA called *quantum logic array* (QLA) which is a 2-D array of super-cells called *tiles* [6]. Each tile comprises of an $n \times n$ array of cells so as a logical operation can fit in. Thaker *et al.* observe that the parallelism in the quantum circuits is very limited [7]. Hence, they suggest the *compressed QLA* (CQLA) which separates the array into two regions: *memory* and *compute*. In the memory region, the qubits which do not participate in any operation at the current time are stored. These qubits absorb less noise and hence require a lighter error correction scheme. In other words, the error correction needs fewer physical ancilla qubits for every physical data qubit (a ratio of 1 to 2). On the other hand, the qubits in the compute region actively participate in the quantum operations. Hence they require a much larger number of ancilla qubits. Since the compute region occupies much smaller area than the memory region, this new architecture helps saving a lot of unnecessary physical ancilla qubits which are used in QLA. Memory region is also further broken down into the cache and the global memory to address the qubit locality issue required by the compute region.

• **Quantum Mapping Techniques.** The quantum mapping problem, similar to the corresponding problem in the traditional VLSI area, is known as a hard problem. Whitney *et al.* suggest a CAD flow for mapping a quantum circuit fault-tolerantly to an ion-trap fabric [2]. To address the scalability issue, they adopt the two-level (physical and logical) mapping. Other levels of hierarchy are handled manually without any automation. Jones *et al.* propose a five-layer stack for implementing a quantum computer [8]. This work does not show how to overcome the complexity of the "*logical layer*" and tries to address other complexities in the design by adding more layers. In [1], we have suggested to use a quick estimation method called *LEQA* to calculate the circuit latency instead of a full-fledged mapping. Even though this approach is quite fast, it does not provide the detailed mapping. Moreover, it requires a flattened high-level netlist as the input which requires a huge amount of disk space to store the netlist and a large memory in order to store its data structures. Additionally, LEQA does not consider the ancilla sharing problem. Although several heuristics have been proposed in the literature for solving the quantum mapping problem, none of them is able to deal with large circuits [2][6][7][9][10].

# 3. PROPOSED ARCHITECTURE

The CQLA architecture reviewed in the previous section assumes that the number of required ancilla qubits for all of the logical operations followed by the QEC is the same. Hence, CQLA accounts for a certain amount of physical ancilla qubits for every logical operation in the compute region. However, this assumption is not true. An important subset of logical operations, called *non-transversal* operations, requires more ancilla than *transversal* operations. It has been proven that every universal logical operation set contains at least one non-transversal gate which varies based on the employed QEC [11]. Table 1 summarizes the ancilla requirements for two typical QEC codes and various logical operations. As can be seen, a non-transversal operation requires half an order of magnitude (in the *Steane code*) up to more than one order of magnitude (in the *Bacon-Shor code*) more ancilla qubits compared to that of transversal operations. Moreover, a two-qubit transversal operation (like CNOT) requires twice ancilla qubits compared to that of a one-qubit transversal operation.

**Table 1. Ancilla requirements for various QEC codes and operations**

| QEC | Operation Type | Operation | # of Ancilla Qubits |
|---|---|---|---|
| **Steane ⟦7,1,3⟧** | Transversal | X, Y, Z, H, S | 28 |
| | | CNOT | 56 |
| | Non-Transversal | T | 100 |
| **Bacon-Shor ⟦9,1,3⟧** | Transversal | X, Y, Z, H | 18 |
| | | CNOT | 36 |
| | Non-Transversal | S | 58 |
| | | T | 309 |

With this observation, the compute region cannot be a pre-allocated area with a fixed number of ancilla qubits for all of operations; otherwise, it leads to an overestimation of the required ancilla. Hence, we propose the *quantum reconfigurable compute region* (QRCR) which distributes the ancilla qubits in the compute region based on the dispatched operations. In other words, the ancilla qubits are *shared* among the operations which are being executed based on their ancilla qubit requirements. To further speed up the computation and eliminate the overhead of qubit routing, a hierarchical memory design is adopted. The first level of the hierarchy is the *quantum cache* which stores qubits that are immediately needed after the execution of the current operations in the QRCR. The second level is the *quantum memory* which keeps the rest of the qubits. Using this hierarchy, the overhead of the routing delay can be mostly hidden. More precisely, the routing delay is substantially smaller than the delay of logical operations, because the routing involves qubit movement (or information swap) which can be done directly by using fast primitive operation(s), whereas logical operations require time consuming QECs. The only considerable routing delay is the time required to load the qubits from the quantum cache to the QRCR.

Figure 3 (a) depicts a *quantum core* which is comprised of a QRCR, a quantum cache, and a quantum memory. As can be seen, QRCR is located at the center and surrounded by the quantum cache followed by the quantum memory. The highly shaded areas inside the QRCR have higher number of ancilla, whereas lightly shaded areas contain lesser ancilla. The arrangement of ancilla changes during the runtime of a quantum algorithm based on the operations being executed.

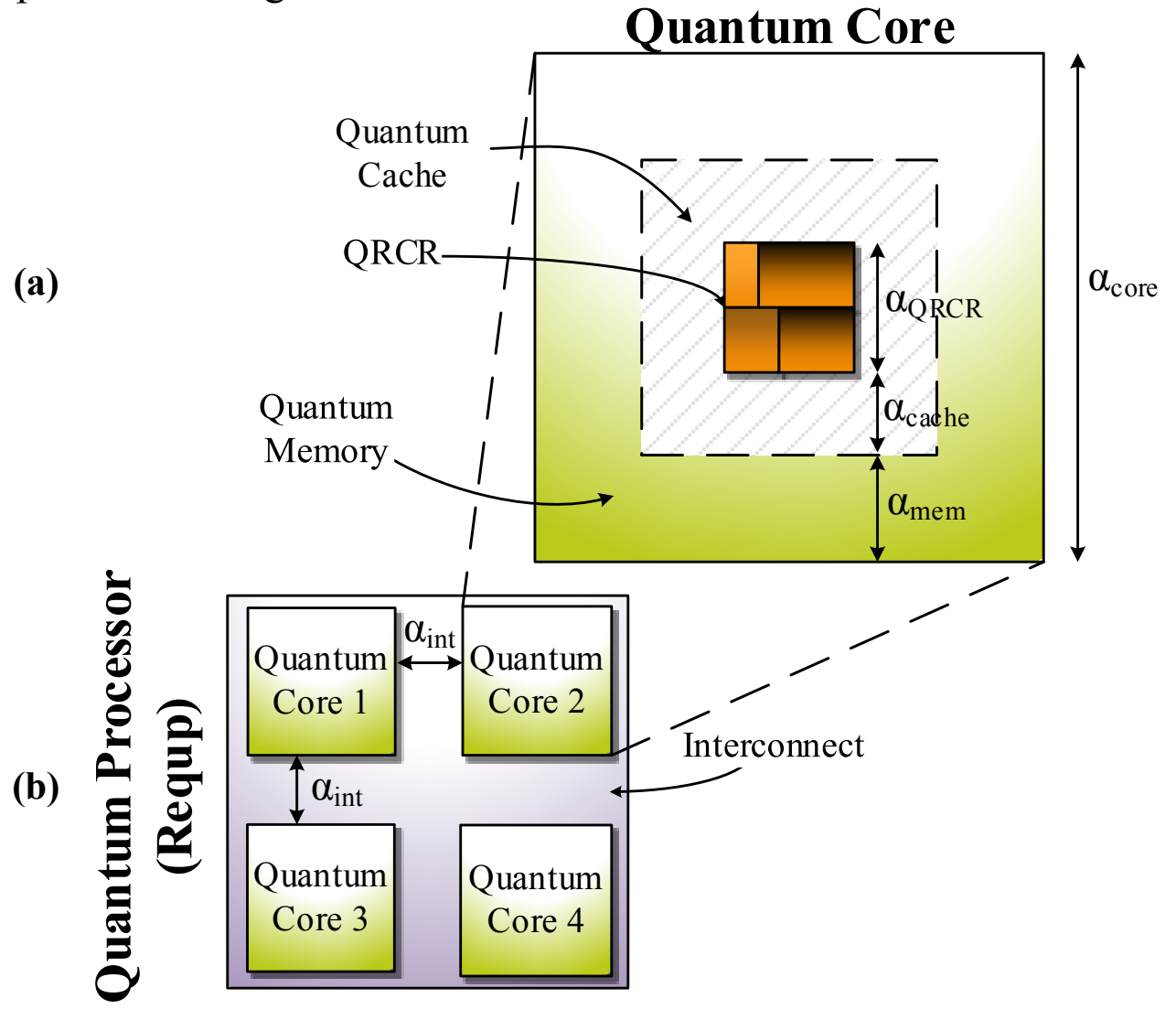

**Figure 3. (a) Structure of a quantum core (b) Structure of a quad-core Requp**

In large-scale algorithms, the size of the cache and the memory may grow. This increases the qubit routing delay which was already hidden by the long delay of logical operations. To avoid this effect, we further extend the quantum core architecture to the *reconfigurable quantum processor architecture* (Requp). A Requp contains multiple reconfigurable quantum cores which are connected to each other by *quantum interconnects*. Quantum interconnects are physically implemented similar to the rest of the quantum physical fabric. Here, this distinction is made for clarity. A quad-core Requp is shown in Figure 3 (b).

# 4. SQUASH

This section introduces the *scalable quantum mapper considering ancilla sharing* (*Squash*). Squash adopts Requp as its underlying fabric abstraction. It takes a netlist of quantum operations in the *quantum assembly* (QASM) format [12], a QEC code description similar to Table 1, the number of quantum cores ($k$), the delay of a qubit travelling the extent of one grid cell (called *qubit unit-distance delay* and denoted by $\beta_{PMD}$), interconnect width ($\alpha_{int}$), a coefficient which models the effect of memory size on the routing speed ($\gamma_{mem}$), and the total ancilla budget ($A$). The quantum operation set is limited to the fault-tolerant operation set. The output of Squash is a circuit mapped to the designated fabric. Algorithm 1 presents the steps involved in Squash.

As it is explained previously, early work found that quantum algorithms offer limited parallelism [7]. By investigating various quantum algorithms, including the phase estimation algorithm which is at the basis of many well-known and useful quantum algorithms (such as Shor's factorization algorithm [3] and quantum random walk [4]), we realized that quantum algorithms can be divided into major *computational stages* which cannot be run in parallel, i.e., they should be executed serially. The main reason is due to the *no-cloning theorem*, which does not allow a qubit to be replicated. This limitation forbids any fan-out in a quantum circuit. As a result, scheduling of computational stages becomes a trivial task— they should be run serially. Moreover, these stages tend to contain a small number of repetitively-used quantum kernels.

**Algorithm 1: Squash**

**Input:** A QASM, a QEC code, Requp parameters (i.e., number of quantum cores ( $k$), qubit unit-distance delay ($\beta_{PMD}$), interconnect width ($\alpha_{int}$), and memory size effect on the routing coefficient ($\gamma_{mem}$)), and total ancilla budget ($A$)
**Output:** Mapped circuit of the given quantum algorithm

1. Identify a set of quantum kernels $\mathcal{S} = \{\mathcal{S}_1, \dots, \mathcal{S}_m\}$
2. **For** each $\mathcal{S}_i$ **in** $\mathcal{S}$
3. Generate a QODG for the operations in $\mathcal{S}_i$ ($\text{QODG}_i$)
4. K-way partition the $\text{QODG}_i$ to get $\mathcal{P}_i = \{\mathcal{P}_{i,1}, \dots, \mathcal{P}_{i,k}\}$
5. Calculate the routing delay matrix $\boldsymbol{d}$
6. Bind each $\mathcal{P}_{i,j}$ to one of the quantum cores
7. Map each $\mathcal{P}_{i,j}$ to the designated quantum core
8. **End For**
9. **Return** mapping of $\{\mathcal{P}_{i,j} | 1 \le i \le m, 1 \le j \le k\}$.

Mapping only one instance of these kernels significantly reduces the runtime overhead. Accordingly, the first line of Algorithm 1 identifies a list of candidates for the quantum kernels. Moreover, in the for-loop block (lines 3 to 7), the algorithm maps each of the kernels separately. Then, the entire mapping solution can be constructed by properly ordering the mapping results for each of the kernels (line 9).

In the rest of this section, the details of mapping a quantum kernel to the given Requp is explained (i.e., the for-loop body). Line 3 generates a QODG as explained in Section 2. Next, it is broken into $k$ parts such that $k$ quantum cores can execute these parts simultaneously while having the minimum amount of inter-core communication (line 4). Next, the routing delay matrix is calculated, which comprises of the qubit routing delay between every pair of quantum cores (line 5). Each part is then bound (line 6), and finally mapped (line 7) to a quantum core.

## 4.1 QODG K-Way Partitioning

A standard $k$-way partitioning algorithm takes a graph, and divides its node set into $k$ disjoint parts such that the parts are balanced in terms of their size and a minimum number of edges are cut. Using this method, the same workload is assigned to each quantum core, while inter-core communication is minimized. However, there is no guarantee that parts can be executed in parallel which is in fact a desired metric in order to reduce the runtime. As an example, consider the QODG shown in Figure 2 (b), and assume a 2-way partitioning is needed. A standard graph partitioning algorithm may suggest the dashed cut which partitions the graph into two parts with almost equal number of nodes. Unfortunately, this solution does not allow any parallelism. On the other hand, the dotted cut, even though one part has twice as many nodes as the other one, is a better solution as parts can be executed simultaneously.

In order to guide the partitioning algorithm to produce parts that can be run in parallel, we employ the technique proposed in [13] which adopts the *multi-constraint graph partitioning* (MCGP) method [14]. The MCGP method assigns an $n_{con}$-dimensional weight vector to each node, and then balances the total sum of the weight values among the parts in each dimension while minimizing the edge cut. The weight vector for each QODG node is calculated as follows. Initially, the QODG is levelized. Let $n_i$ be the number of nodes at level $i$ ($L_i$), $U = \{L_i \,| n_i \ge k\}$, and $n_{con} = |U|$ (i.e., the number of levels that contain more than $k-1$ nodes.) Then, weight vectors of size $n_{con}$ are assigned to each node. For nodes that are at level $L_i \notin U$, the weight vector is set to zero vector. For other nodes, we first assign a label to each level $L_i \in U$ using the one-hot coding scheme. This label will be used as the weight vector for all of the nodes within the same level. Hence, by using one-hot coding, a unique dimension of the weight vector is assigned to all nodes at level $L_i \in U$. Therefore, the MCGP method is forced to partition these nodes into distinct parts so that the total weight in the corresponding dimension for each part is balanced. An example for the 2-way MCGP is shown in Figure 2 (b).

## 4.2 Routing Delay Matrix Calculation

In this phase, based on the information obtained from the partitioning step, the quantum core is characterized in order to find the accurate qubit routing delays between each pair of cores. Note that it is not necessary to use the same quantum core configuration for all of the quantum kernels, because it is just an abstraction to simplify the mapping and hide the technology details. For this purpose, four parameters, namely $\alpha_{QRCR}$, $\alpha_{core}$, $\alpha_{cache}$, and $\alpha_{mem}$ (which are shown in Figure 3) are initially calculated. The approach is to derive the number of physical qubits each area should accommodate and then the desired distances are calculated accordingly. $\alpha_{QRCR}$ can be obtained by

$$\alpha_{QRCR} = \left\lceil \sqrt{\frac{A/k}{A_{min}} . L_{code} + \left(A/k - \frac{D_{max}}{2}\right)} \right\rceil, \quad (1)$$

where $A_{min}$ is the minimum ancilla qubit requirement among quantum operations, $L_{code}$ is the QEC code length, and $D_{max}$ is the maximum number of data qubits a core may accommodate. For instance, for the Steane code listed in Table 1, $A_{min} = 28$ and $L_{code} = 7$. $D_{max}$ can be calculated by referring to the partitioned set of operations for each core. The first summation term in Equation (1) accounts for the maximum number of physical data qubits the QRCR may host, whereas the second term accounts for the physical ancilla qubits. Note that $A/k$ is the ancilla budget per core. Furthermore, $\frac{D_{max}}{2}$ ancilla qubits are reserved for the error correction of data qubits in the cache and the memory. As mentioned earlier, for the QEC of every two data qubits in the cache or the memory, one ancilla qubit is enough. $\alpha_{core}$ is determined by

$$\alpha_{core} = \left\lceil \sqrt{D_{max} . L_{code} + A/k} \right\rceil. \quad (2)$$

As suggested in [7], $\alpha_{cache}$ can be set such that the cache area becomes twice as large as the QRCR area. Hence, $\alpha_{cache}$ can be calculated as

$$\alpha_{cache} = min\left\{\left\lceil \frac{\sqrt{3}-1}{2} \alpha_{QRCR} \right\rceil, \frac{\alpha_{core} - \alpha_{QRCR}}{2}\right\}. \quad (3)$$

A minimum value is calculated in order to avoid over provisioning of resources for the cache, i.e., the cache plus QRCR area should not be larger than the area of the core. Finally, $\alpha_{mem}$ can be derived based on the values of $\alpha_{QRCR}$, $\alpha_{cache}$, and $\alpha_{core}$:

$$\alpha_{mem} = \left\lceil \frac{\alpha_{core}}{2} - \frac{\alpha_{QRCR}}{2} - \alpha_{cache} \right\rceil. \quad (4)$$

Using these four parameters, the communication delay for routing a qubit from the QRCR of core $x$ to the QRCR of core $y$ can be calculated as

$$d_{x,y} = \begin{cases} n_{x,y}(\alpha_{core} + \alpha_{int})\beta_{PMD}, & x \ne y \\ \dfrac{(\alpha_{QRCR} + \alpha_{cache} + \gamma_{mem}\alpha_{mem})}{2} \beta_{PMD}, & x = y \end{cases} \quad (5)$$

where $n_{x,y}$ is the Manhattan distance between core $x$ and core $y$. The first case ($x \ne y$) is considered as the inter-core routing delay, whereas the second case ($x = y$) accounts for the delay of transferring a qubit from the cache (or possibly the memory) into the QRCR. Coefficient $\gamma_{mem}$ ensures the proper contribution of the memory size to the routing delay of a qubit. In other words, if the memory size becomes large enough, then the routing delay cannot be overshadowed by the long operation delay, and hence should be considered in the routing delay calculation. We capture this effect with the $\gamma_{mem}$ coefficient.

## 4.3 Resource Binding

After partitioning the QODG, the resultant parts should be bound to the quantum cores such that the total routing delay of qubits between cores is minimized. Since the scheduling of the QODG is not known at this step, we cannot focus on minimizing the total routing delay of the operations on the critical path. Furthermore, the scheduling requires this binding information in order to properly schedule two dependent operations assigned to two different quantum cores.

The binding problem can be formulated as follows:

$$\min \sum_{m=1}^{k} \sum_{n=1}^{k} \sum_{x=1}^{k} \sum_{y=1}^{k} a_{m,n} a_{x,y} d_{n,y} w_{m,x} \quad (6)$$

subject to

$$\sum_{n=1}^{k} a_{m,n} = 1, \text{ for } 1 \le m \le k, \quad (7)$$

$$\sum_{m=1}^{k} a_{m,n} = 1, \text{ for } 1 \le n \le k, \quad (8)$$

where $a_{m,n}$ is a binary variable, which is 1 if $\mathcal{P}_{i,m}$ is bound to quantum core $n$ and 0 otherwise, and $w_{m,x}$ denotes the number of qubits that traverse from part $\mathcal{P}_{i,m}$ to $\mathcal{P}_{i,x}$. The objective function (6) is the sum of inter-core communication delays while constraints (7) and (8) ensure a one-to-one assignment between parts and quantum cores. Since $k$ is fairly small, the computation time to solve the resulting *0-1 quadratic program* (0-1 QP) is of little concern.

## 4.4 Mapping

The objective of scheduling the QODG on $k$ quantum cores is to minimize the overall latency while ensuring that the number of ancilla qubits used in each quantum core is no more than the given budget. The aforesaid scheduling problem is similar to the well-known *minimum-latency resource-constraint multi-cycle* (MLRC-MC) scheduling problem [15] in high-level synthesis with the following difference. The MLRC-MC problem does not deal with the cost of moving data among resources whereas in our formulation the resources (i.e., quantum cores) lie on a given grid, and therefore, their average communication costs can be pre-calculated (see Equation (5)). More precisely, our problem formulation is as follows.

$$\min \mathcal{L} \quad (9)$$

subject to

$$\sum_{O_x \in \mathcal{P}_{i,j}} \sum_{y=0}^{T_x-1} b_{x,z-y} A_{O_x} \le A/k, \ 1 \le z \le \mathcal{L}_{init}, 1 \le j \le k \quad (10)$$

$$\sum_{j=1}^{\mathcal{L}_{init}} b_{x,y} = 1, \forall O_x, \quad (11)$$

$$S_x + T_x + d_{m,n} \le S_y, O_x \to O_y, O_x \in C_m \text{ and } O_y \in C_n, \quad (12)$$

$$S_x + T_x - 1 \le \mathcal{L}, \forall O_x \text{ without any successors}, \quad (13)$$

where $\mathcal{L}$ is the total number of scheduling levels, $T_x$ is the delay of operation $O_x$, $b_{x,y}$ is a binary variable which is 1 if $O_x$ is scheduled to start at scheduling level $y$ and 0 otherwise, $A_{O_x}$ denotes the ancilla requirement of operation $O_x$, $\mathcal{L}_{init}$ is an upper bound for the total number of scheduling levels ($\mathcal{L}$), $S_x$ is equal to the scheduling level where $O_x$ is scheduled, i.e., $b_{x,S_x} = 1$, and $O_x \in C_m$ means that operation $O_x$ is bound to quantum core $m$. Equation (10) sets a constraint on the total number of ancilla that each core can use at each scheduling level. Equation (11) ensures that all of the operations are scheduled. Equation (12) makes sure that dependent operations are properly scheduled, i.e., an operation starts after its predecessor in the QODG is finished. Equation (13) assures that the operations in the last scheduling level are scheduled to finish their execution before or at the scheduling level $\mathcal{L}$. We modified the list scheduling method presented in [16] as described above to solve the scheduling problem.

Using the Requp architecture, the ancilla sharing problem is solved during the scheduling. Moreover, the placement problem has already been solved in the prior step (i.e., resource binding step). Additionally, as it is explained earlier, the routing delay is hidden by the operation delay. Hence, a simple routing algorithm like the xy-routing fits well for the purpose of transferring qubits (or equivalently swapping their information) through the interconnection network of Requp.

# 5. EXPERIMENTAL RESULTS

Squash is developed in Java. It uses METIS 5 [14] as the partitioning engine and Gurobi 5.6 [17] for solving the 0-1 QP. A computer with an Intel Core i7-3770 CPU running at 3.40 GHz and 8GB of memory is employed for the experiments.

The ⟦7,1,3⟧ Steane code with the information presented in Table 1 is adopted as the QEC code. Moreover, the input variables of Squash are set as follows: $\beta_{PMD}$=10 $\mu s$, $\alpha_{int} = 3$, and $\gamma_{mem} = 0.2$. Squash is not limited to a particular quantum technology; however, the ion-trap technology is selected since it is the most promising method for realizing quantum circuits to date [18]. The delay of quantum operations in this technology is taken from [19].

In the rest of this section, first the latency-ancilla count trade-off in quantum circuits is studied using Squash. Then the optimum number of quantum cores for a representative benchmark is found. After that, the resource requirement of Requp, CQLA, and QLA are analytically compared. Finally, Squash is compared with the state-of-the-art mapper.

- **Investigating the latency-ancilla count trade-off:** As it is explained earlier, ancilla qubits are precious resources in quantum computers. Increasing the total ancilla budget lowers the circuit latency and vice versa. In order to study this effect using Squash, the *binary welded tree* (BWT) algorithm [20] is selected as the benchmark and compiled with Scaffold Compiler (which is introduced in [21]) to produce a QASM file. The number of quantum cores ($k$) is set to 4. The trade-off between latency and the ancilla budget ($A$) is shown as a Pareto curve in Figure 4. As can be seen, the delay value saturates at $A = 800$. This means that the circuit does not require more than 800 ancilla qubits.

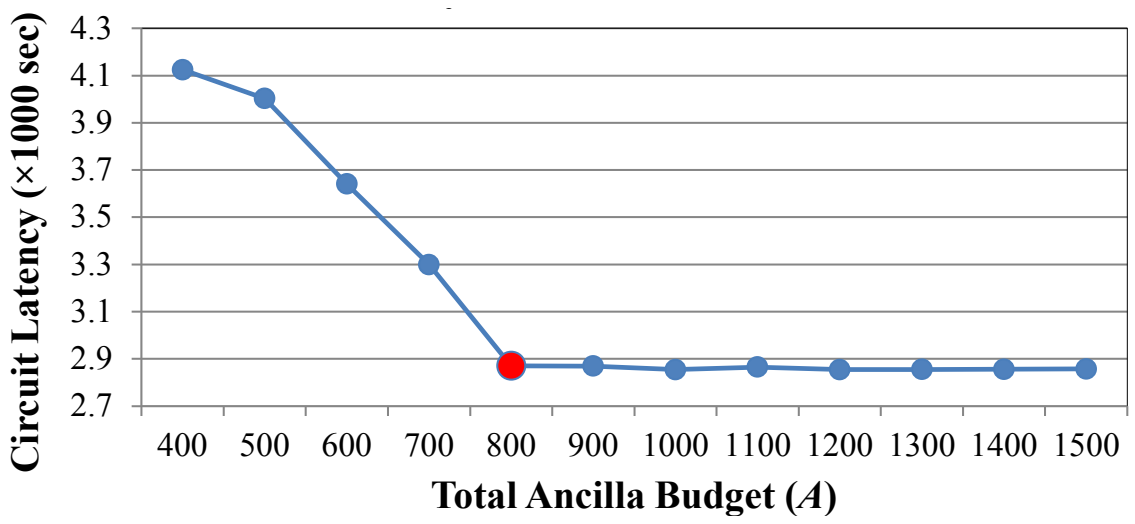

**Figure 4. Latency-ancilla count trade-off for the BWT algorithm using a quad-core Requp architecture**

- **Finding the optimum number of quantum cores:** One of the Squash input parameters is the quantum core count ($k$). The optimum value for this parameter varies based on the parallelism inside a given circuit. The quantum core count is just an abstraction and has no relation with the usage of quantum physical resources. However, Squash has a higher runtime for smaller values of $k$, because the size of the weight vector for partitioning is larger. Figure 5 shows the latency of the BWT algorithm as a function of quantum core count when $A = 800$. It can be seen that the optimal latency is achieved when $k$ is set to 2. However, $k = 4$ is preferred since the runtime of Squash for this case is 15 times faster than that of the former case.
- **Resource usage comparison among Requp, CQLA, and QLA architectures:** In the QLA architecture, every qubit requires to be placed in a quantum tile. Each tile needs to support all types of quantum operations and their respective QEC codes. Hence, in the case of one-level ⟦7,1,3⟧ Steane code, the required ancilla in this architecture is equal to 100×(total qubit count). CQLA limits this value to the maximum number of parallel operations the architecture should be able to execute. For instance, if $z$ parallel operations are supported (which is significantly smaller than the total qubit count), 100×$z$ ancilla qubits are required. Requp improves this resource limitation by considering the fact that all of the parallel operations may not require the maximum number of ancilla qubits (i.e., 100). Therefore, Requp allows to run at most (100/28)$\times z$ operations at the same time while still having the same worst case parallelism as CQLA. This discussion reveals that Requp performs more efficiently in the average case compared to CQLA and behaves as bad as CQLA in the worst case.

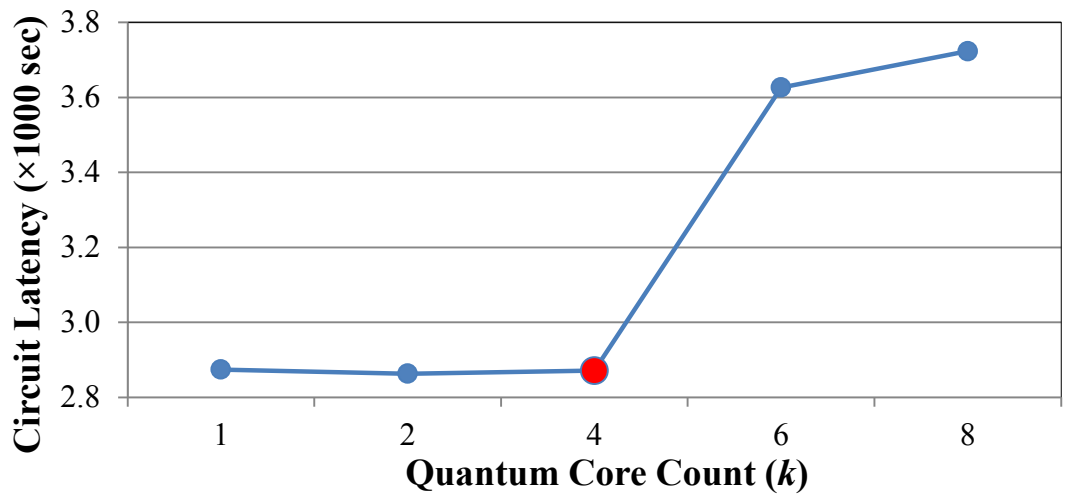

**Figure 5. Finding the optimum number of quantum cores for the BWT algorithm**

- **Comparison between Squash and QSPR:** In this section, the performance and the quality of results produced by Squash is compared with that of *QSPR* which is introduced in [10]. QSPR is a full-fledged quantum mapper which is recently improved to support the QLA architecture [1]. Unfortunately, no quantum mapper for the CQLA architecture is available for the comparison.

Various sizes of the BWT algorithm is compiled based on a parameter called $s$. This parameter is varied from 3 to 19, where $s = 19$ is the problem size of interest. (For the previous experiments, $s$ was set to 5.) Figure 6 compares the circuit latency mapped by Squash and QSPR. As can be seen, Squash could achieve better results in all of problem sizes. This is due to the improved qubit routing mechanism in Squash. As it was explained earlier, Squash hides most of the routing delay by parallelizing it with the execution of logical operations.

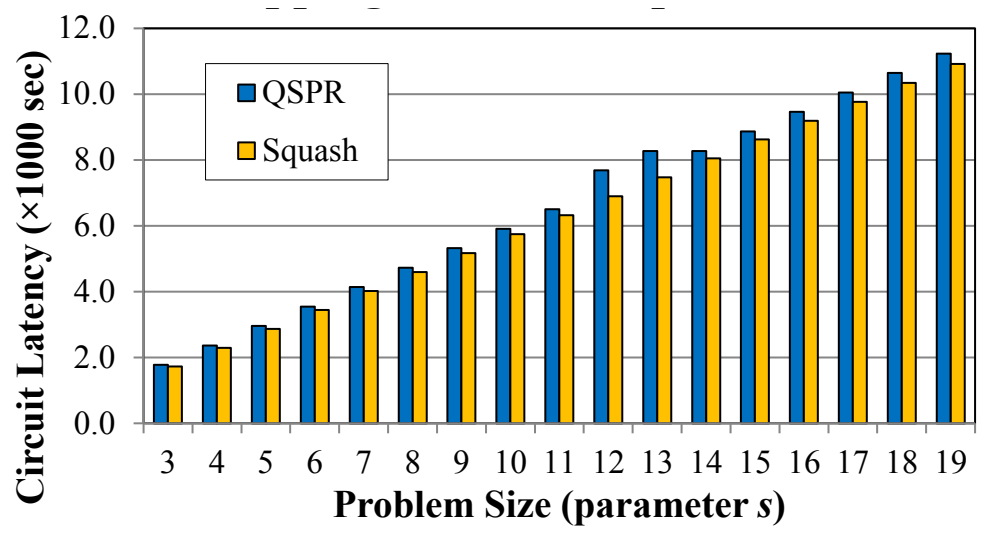

**Figure 6. Comparison of the circuit latency mapped by QSPR and Squash**

Figure 7 compares the runtime of QSPR and Squash. As can be seen, for very small problem sizes ($s < 8$), QSPR is slightly faster than Squash. However, as the problem size grows, the runtime of QSPR radically increases, whereas the runtime of Squash does not change. This phenomenon is due to the fact that QSPR handles a large netlist of quantum operations, whereas Squash maps only the quantum kernels which grow very slowly compared to the main circuit size. Also note that when $s > 15$, QSPR runtime rapidly grows due to the inefficient handling of large netlists.

# 6. CONCLUSION

Quantum circuits for solving real-size problems are gigantic. As a result, quantum mappers have difficulty in mapping them to quantum fabrics. Moreover, current quantum mappers cannot properly handle the ancilla sharing problem which allows reducing the resource demand (even though it is achieved at the cost of increased latency). To address these two key problems, a scalable quantum mapper, called Squash, was introduced. Squash uses a novel multi-core reconfigurable quantum processor architecture, called Requp, which supports a layered approach to mapping a quantum algorithm and enables ancilla sharing. Experimental results demonstrated that Squash can handle large-scale quantum algorithms while providing an effective mechanism for sharing ancilla qubits.

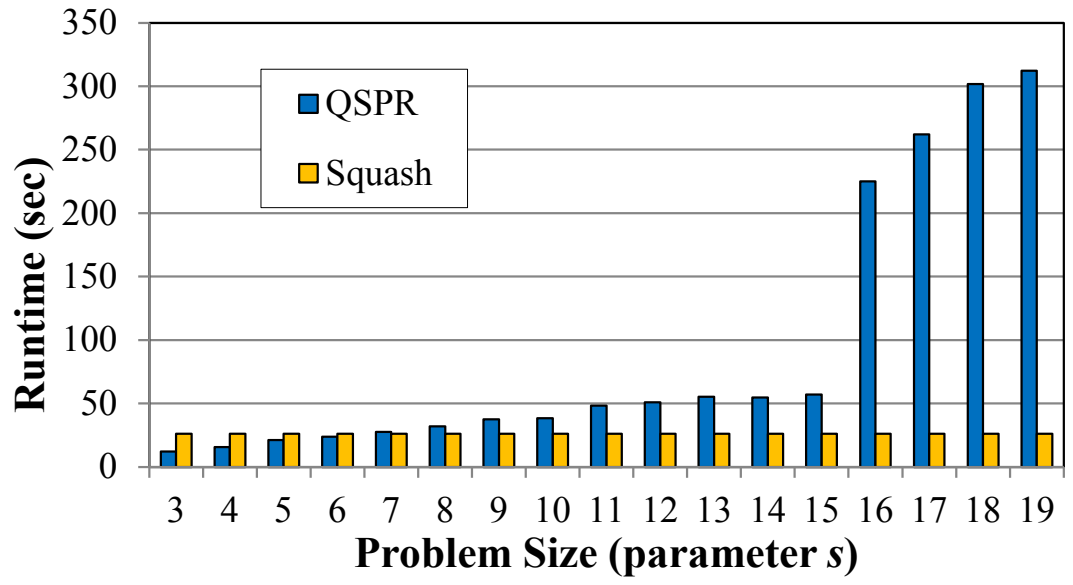

**Figure 7. Comparison of mapping results between QSPR and Squash**

# 7. ACKNOWLEDGMENTS

The authors would like to thank Professor Todd Brun for his valuable comments about the calculation of ancilla requirements for various QEC codes and operations.

This research was supported in part by the Intelligence Advanced Research Projects Activity (IARPA) via Department of Interior National Business Center contract number D11PC20165.

# 8. REFERENCES

[1] M. J. Dousti and M. Pedram, "LEQA: Latency Estimation for a Quantum Algorithm Mapped to a Quantum Circuit Fabric," in *DAC*, 2013.

[2] M. G. Whitney et al., "A Fault Tolerant, Area Efficient Architecture for Shor's Factoring Algorithm," in *ISCA*, 2009.

[3] M. A. Nielsen and I. L. Chuang, *Quantum Computation and Quantum Information*. Cambridge University Press, 2010.

[4] S. E. Venegas-Andraca, *Quantum walks for computer scientists*. Morgan & Claypool Publishers, 2008.

[5] V. V. Shende and I. L. Markov, "On the CNOT-cost of TOFFOLI gates," *QIC*, vol. 9, no. 5, pp. 461–486, 2009.

[6] T. S. Metodi et al., "A Quantum Logic Array Microarchitecture: Scalable Quantum Data Movement and Computation," in *MICRO*, 2005.

[7] D. D. Thaker et al., "Quantum Memory Hierarchies: Efficient Designs to Match Available Parallelism in Quantum Computing," in *ISCA*, 2006.

[8] N. C. Jones et al., "Layered Architecture for Quantum Computing," *Phys. Rev. X*, vol. 2, no. 3, p. 031007, 2012.

[9] L. Kreger-Stickles and M. Oskin, "Microcoded Architectures for Ion-Tap Quantum Computers," in *ISCA*, 2008.

[10] M. J. Dousti and M. Pedram, "Minimizing the Latency of Quantum Circuits During Mapping to the Ion-Trap Circuit Fabric," in *DATE*, 2012.

[11] B. Zeng et al., "Transversality Versus Universality for Additive Quantum Codes," *IEEE Trans. Inf. Theory*, vol. 57, no. 9, pp. 6272–6284, 2011.

[12] "Quantum Architectures: qasm2circ." [Online]. Available: http://www.media.mit.edu/quanta/qasm2circ.

[13] M. Tanaka and O. Tatebe, "Workflow Scheduling to Minimize Data Movement Using Multi-constraint Graph Partitioning," in *CCGRID*, 2012.

[14] G. Karypis and V. Kumar, "Multilevel Algorithms for Multi-constraint Graph Partitioning," in *Supercomputing*, 1998.

[15] C.-T. Hwang et al., "A Formal Approach to the Scheduling Problem in High Level Synthesis," *IEEE TCAD*, vol. 10, no. 4, pp. 464–475, 1991.

[16] G. D. Micheli, *Synthesis and Optimization of Digital Circuits*. McGraw-Hill, 1994.

[17] "Gurobi Optimizer." [Online]. Available: http://www.gurobi.com.

[18] T. D. Ladd et al., "Quantum computers," *Nature*, vol. 464, no. 7285, pp. 45–53, 2010.

[19] H. Goudarzi et al., "Design of a Universal Logic Block for Fault-Tolerant Realization of any Logic Operation in Trapped-Ion Quantum Circuits," *Quantum Inf. Process.*, pp. 1–33, Jan. 2014.

[20] A. M. Childs et al., "Exponential Algorithmic Speedup by a Quantum Walk," in *Proceedings of the Theory of Computing*, 2003.

[21] A. JavadiAbhari et al., "ScaffCC: A Framework for Compilation and Analysis of Quantum Computing Programs," in *CF*, 2014.